\documentclass[aps, twocolumn, showpacs, amsmath]{revtex4-2}
\usepackage{dcolumn}
\usepackage{bm}
\usepackage{graphicx}
\usepackage{color}
\usepackage{amsthm}
\usepackage{amssymb}
\usepackage{hyperref}

\begin{document}
\def\be{\begin{equation}}
\def\ee{\end{equation}}

\def\bc{\begin{center}}
\def\ec{\end{center}}
\def\bea{\begin{equation}}
\def\eea{\end{equation}}
\newcommand{\avg}[1]{\langle{#1}\rangle}
\newcommand{\Avg}[1]{\left\langle{#1}\right\rangle}

\def\ie{\textit{i.e.}}
\def\etal{\textit{et al.}}
\def\m{\vec{m}}
\def\G{\mathcal{G}}

\newcommand{\davide}[1]{{\bf\color{blue}#1}}
\newcommand{\gin}[1]{{\bf\color{green}#1}}

\title{Random walks on complex networks under node-dependent stochastic resetting}

\author{Yanfei Ye}
\author{Hanshuang Chen}\email{chenhshf@ahu.edu.cn}

\affiliation{School of Physics and Optoelectronics Engineering, Anhui
	University, Hefei 230601, China}

\begin{abstract}
In the present work, we study random walks on complex networks subject to stochastic resetting when the resetting probability is node-dependent. Using a renewal approach, we derive the exact expressions of the stationary occupation probabilities of the walker on each node and the mean first passage time between arbitrary two nodes. Finally, we demonstrate our theoretical results on three networks with two different resetting protocols, validated by numerical simulations as well. We find that under a delicate setting it is advantageous to optimize the efficiency of a global search on such networks by the node-dependent resetting probability. 
\end{abstract}

\maketitle


\section{Introduction}\label{sec1}
First passage underlies a wide variety of stochastic phenomena
across diverse fields \cite{redner2001guide,van1992stochastic,klafter2011first,bray2013persistence}. Indeed, chemical
and biochemical reactions \cite{RevModPhys.85.135}, foraging strategies of
animals \cite{RevModPhys.83.81}, and the spread of diseases on social networks
or of viruses through the world wide web \cite{RevModPhys.87.925} are
often controlled by first encounter events.

In the last decade, there has been an increasing interest in first passage under resetting (see \cite{evans2020stochastic}  for a recent review). Resetting refers to a sudden interruption of a stochastic process followed by its starting
anew. Interestingly, for a one-dimensional Brownian motion subject to stochastic resetting \cite{evans2011diffusion}, the occupation probability at stationarity is strongly altered. The mean time to reach a given target for the first time can become finite and be minimized with respect to the resetting rate. Some other interesting features of resetting Brownian motions or random walks have also been unveiled. The mean perimeter and the mean area of the convex hull of a two-dimensional resetting Brownian motion were exactly computed, which showed the two quantities grow much slowly with time than the case without resetting \cite{PhysRevE.103.022135}. For random walks on a $d$-dimensional hypercubic lattice under resetting \cite{arXiv:2202.04906}, the average number of distinct sites visited by the walker grows extremely slowly with the time steps, and the so-called recurrence-transience transition at $d=2$ for standard random walks (without resetting) disappears in the presence of resetting. In a finite one-dimensional domain, the distribution of the number of distinct sites visited by a random walker before hitting a target site with and without resetting was deduced, and the distribution can be simply expressed in terms of splitting probabilities only \cite{PhysRevE.103.032107}.
Moreover, different types of resetting protocols and Brownian motions have been
considered, such as temporally or spatially dependent resetting rate \cite{evans2011diffusion2,pal2016diffusion,Pinsky2020,PhysRevE.96.022130}, in the presence of external potential \cite{pal2015diffusion,ahmad2019first,gupta2020stochastic}, run-and-tumble particles \cite{evans2018run,santra2020run,bressloff2020occupation}, active particles \cite{scacchi2018mean,kumar2020active}, and so on \cite{basu2019symmetric}.  
These studies have triggered an enormous recent activities in the field, including statistical physics \cite{pal2017first,gupta2014fluctuating,evans2014diffusion,meylahn2015large,PhysRevLett.116.170601,chechkin2018random,rose2018spectral,magoni2020ising,de2020optimization}, stochastic thermodynamics \cite{fuchs2016stochastic,pal2017integral,gupta2020work}, chemical and biological
processes \cite{reuveni2014role,rotbart2015michaelis}, optimal control theory \cite{arXiv:2112.11416}, and single-particle experiments \cite{tal2020experimental,besga2020optimal}. 

Random walks on complex networks not only underlie many important stochastic dynamical processes on networked systems \cite{PhysRevLett.92.118701,PhysRevE.87.012112,zhang2011mean,masuda2017random}, such as transmission of virus or rumors \cite{RevModPhys.87.925,colizza2007reaction,PhysRevX.1.011001}, population extinction \cite{WKBReview1,PhysRevLett.117.028302}, neuronal firing \cite{tuckwell1988introduction}, consensus formation \cite{PhysRevLett.94.178701}, 
but also find a broad range of applications, such as community detection \cite{rosvall2008maps,zhou2004network,pons2005computing}, human mobility \cite{PhysRevE.86.066116,riascos2017emergence,barbosa2018human}, ranking and searching on the web \cite{PhysRevLett.92.118701,newman2005measure,lu2016vital,kleinberg2006complex,RevModPhys.87.1261}. However, the impact of resetting on random walks in networked systems has  received only a small amount of attention \cite{avrachenkov2014personalized,avrachenkov2018hitting,christophorov2020peculiarities,PhysRevE.103.012122,PhysRevE.103.052129,huang2021random}. Random walks on networks under resetting have many applications in computer science and physics. For instance, label propagation in machine learning algorithms \cite{Bautista2019}, or the famous PageRank \cite{Pagerank1998}, can be interpreted
as a random walker with uniform resetting probability to all the nodes of the network.
Human and animal mobility consists of a mixture of short-range moves with intermittent long-range moves where an agent relocates to a new place and then starts local moves \cite{Barabasi.Nature2008,Walsh_NatPhys2010,RevModPhys.83.81}.
Until recently, Riascos \textit{et al.} studied the impact of stochastic resetting with a constant probability on random walks on arbitrary networks \cite{PhysRevE.101.062147}. They have established the relationships between the random walk
dynamics and the spectral representation of the transition matrix in the absence of resetting. Furthermore, they discussed the condition under which resetting becomes advantageous to reduce the mean first passage time (MFPT) \cite{arXiv:2110.15437}. 
Subsequently, the result has been generalized to the case when multiple resetting nodes exist \cite{PhysRevE.103.062126,Chaos2021_31.093135}.

In the present work, we aim to generalize the previous study to the case when the resetting probability at each node is not a constant, but is node-dependent. The natural generalization not only brings some new challenges from a theoretical point of view, but also may find practical perspectives in technical aspects. In search processes on networks, if a searcher has partial information about the present position such as node's degree, can one design a node-dependent resetting strategy to enhance the search efficiency? This may be important for heterogeneous networks, encountering on most empirical systems \cite{masuda2017random,PhysRevE.64.046135}.  In the standard teleportation scheme of PageRank, one teleports to nodes uniformly at random, i.e., the probability to land on each node is the same. An alternative choice is a  ``personalized PageRank", in which the landing probability is localized around one node or a small number of nodes \cite{Gleich_SIAMRev_2015}. Such a choice has been shown to be beneficial to reducing the effect of teleportation \cite{PhysRevE.85.056107}, also finding its applications in community detection \cite{PhysRevE.91.012821}. Taking the advantage of renewal structure in Markovian processes, we derive the occupation probability of the walker at each node at stationarity and the MPFT between arbitrary two nodes. We find that the two quantities can be calculated from the matrix defined in Eq.(\ref{eq12}). We then apply our theoretical results to three concrete networks, and consider two different settings of node-dependent resetting probability, i.e, that depends on the shortest path length to the resetting node or node's degree. We observe that both the two settings can further optimize the efficiency of a global search compared with the case when the resetting probability is a constant.   

\section{Model}\label{sec2}
First of all, we define the standard discrete-time random walks on an undirected and unweighted network of size $N$ \cite{PhysRevLett.92.118701}. Assuming that a particle is located at node $i$ at time $t$, at the next time $t+1$ it hops to one of neighboring nodes of node $i$ with equal probability. Thus, the transition matrix $\bf{W}$ among nodes can be written as $\bf{W}=\bf{D}^{-1} \bf{A}$, where $\bf{A}$ is the adjacency matrix of the underlying network, and ${\bf{D}}={\rm{diag}}\{d_1, \cdots, d_N  \}$ is a diagonal matrix with $d_i=\sum_{j=1}^{N} A_{ij}$ being the degree of node $i$. 

We now incorporate stochastic resetting with a node-dependent resetting probability into the standard random walk model. We first choose a node as the only resetting node, labelled with $r$. Then, at each time step, the particle either performs a standard random walk with the probability $1-\gamma_i$ or is reset to the resetting node $r$ with the probability $\gamma_i$. The resetting probability $\gamma_i$ is dependent on some attribute of node $i$. In the following, we consider that $\gamma_i$ is a function of the degree of node $i$ or the shortest path length between node $i$ and the resetting node $r$, although our next deduction is general and can be also applied to other types of functions.

\section{Stationary occupation probability}\label{sec3}

Let us denote by ${P_{ij}}(t)$ the probability that node $j$ is visited at time $t$, providing that the particle has started from node $i$ at $t=0$, which satisfies a first renewal equation \cite{pal2016diffusion,chechkin2018random,PhysRevE.96.022130},  
\begin{equation}\label{eq1}
{P_{ij}}\left( t \right) = P_{ij}^{\rm{\rm{nores}}}\left( t \right) + \sum\limits_{t' = 1}^t {\sum\limits_{k = 1}^N {{\gamma _k}P_{ik}^{\rm{nores}}\left( {t' - 1} \right){P_{rj}}\left( {t - t'} \right)} } 
\end{equation}
where $P_{ij}^{\rm{nores}}(t)$ denote the probability of all possible trajectories that the particle starts from node $i$ at $t=0$ and ends at node $j$ at time $t$, without undergoing any reset event during the time interval $[0, t]$. Therefore, the first term in Eq.(\ref{eq1}) accounts for the particle is never reset up to time $t$, while the second term in Eq.(\ref{eq1}) accounts for the particle is reset at time $t'$ for the first time, after which the process starts anew from the resetting node
for the remaining time $t-t'$.

$P_{ij}^{\rm{nores}}(t)$ can be calculated as 
\begin{equation}\label{eq2}
P_{ij}^{\rm{nores}}(t)=\langle {i} |{{\bf{\tilde W}}}^t| j \rangle,
\end{equation}
where  
\begin{equation}\label{eq3}
\bf{\tilde W}=(\bf{I}-\bf{Y})  \bf{W}.
\end{equation}
Here $|i \rangle$ denotes the canonical base with all its components equal
to 0 except the $i$th one, which is equal to 1. $\bf{I}$ and $\bf{W}$ are respectively the identity matrix and transition matrix without resetting, and ${\bf{Y}}={\rm{diag}}\{ \gamma_1, \cdots, \gamma_N \} $ being a diagonal matrix.  It can be proved that $\bf{\tilde W}$ can be written as the spectral decomposition (see Appendix \ref{app1} for details), ${\bf{\tilde W}}= \sum\nolimits_{\ell = 1}^{N} {{\lambda_\ell}| {{\psi _\ell}} \rangle } \langle {{{\bar \psi }_\ell}} |$, where $\lambda_\ell$ is the $\ell$th eigenvalue of $\bf{\tilde W}$, and the corresponding left eigenvector and right eigenvector are respectively $\langle \bar \psi_\ell$ and $|\psi_\ell \rangle$, satisfying $\langle \bar \psi_\ell | \psi_m \rangle =\delta_{\ell m}$, and $\sum_{\ell=1}^{N} |\psi_\ell \rangle \langle \bar \psi_\ell |  =\bf{I}$. Thus, Eq.(\ref{eq2}) can be rewritten as
\begin{equation}\label{eq4}
P_{ij}^{\rm{nores}}(t)= \sum\limits_{\ell = 1}^{N} {\lambda_\ell^t} \langle i | {\psi _\ell} \rangle  \langle {{{\bar \psi }_\ell}} | j \rangle.
\end{equation}

Performing the Laplace transform for Eq.(\ref{eq1}), $\tilde{f}(s)=\sum_{t=0}^{\infty} f(t) e^{-st}$, which yields
\begin{equation}\label{eq5}
{\tilde P_{ij}}\left( s \right) = \tilde P_{ij}^{\rm{nores}}(s) + {e^{ - s}}{\tilde P_{rj}}(s)\sum\limits_{k=1}^N {{\gamma _k}\tilde P_{ik}^{\rm{nores}}(s)} ,
\end{equation}
where $\tilde P_{ij}^{\rm{nores}}(s)$ can be obtained from Eq.(\ref{eq4}), given by
\begin{equation}\label{eq5.1}
\tilde P_{ij}^{\rm{nores}}(s) = \sum\limits_{\ell = 1}^N {\frac{{\langle i | {{\psi _\ell}} \rangle \langle {{{\bar \psi }_\ell}} | j \rangle }}{{1 - {\lambda _\ell} {e^{ - s}} }}} .
\end{equation}

Letting $i=r$ in Eq.(\ref{eq5}), we obtain
\begin{equation}\label{eq6}
{\tilde P_{rj}}(s) = \frac{{\tilde P_{rj}^{\rm{nores}}(s)}}{{1 - {e^{ - s}}\sum\nolimits_{k=1}^N {{\gamma _{k}}\tilde P_{rk}^{\rm{nores}}(s)} }}.
\end{equation}
Substituting Eq.(\ref{eq6}) into Eq.(\ref{eq5}), we have
\begin{equation}\label{eq7}
{\tilde P_{ij}}(s) = \tilde P_{ij}^{\rm{nores}}(s) + \frac{{{e^{ - s}}\sum\nolimits_{k=1}^N {{\gamma _k}\tilde P_{ik}^{\rm{nores}}\left( s \right)} }}{{1 - {e^{ - s}}\sum\nolimits_{k=1}^N {{\gamma _k}\tilde P_{rk}^{\rm{nores}}(s)} }}\tilde P_{rj}^{\rm{nores}}(s).
\end{equation}


Inverting Eq.(\ref{eq7}) is difficult; however, we can instead calculate the stationary occupation probability by evaluating the limit, 
\begin{equation}\label{eq9}
{P_{j}}(\infty) =  \mathop {\lim }\limits_{s \to 0} \left( {1 - {e^{ - s}}} \right){\tilde P_{ij}}\left( s \right) .
\end{equation}
Substituting Eq.(\ref{eq7}) into Eq.(\ref{eq9}), and after some tedious calculations, we obtain (see Appendix \ref{app2} for details)
\begin{equation}\label{eq10}
{P_j}( \infty  ) = \frac{{\sum\nolimits_{\ell = 1}^N {\frac{{\langle r | {{\psi _\ell}} \rangle \langle {{{\bar \psi }_\ell}} | j \rangle }}{{1 - {\lambda _\ell}}}} }}{{\sum\nolimits_{k = 1}^N  \sum\nolimits_{\ell = 1}^N {\frac{{\langle r | {{\psi _\ell}} \rangle \langle {{{\bar \psi }_\ell}} | k \rangle }}{{{{ {1 - {\lambda _\ell}} }}}}} }}.
\end{equation}
Eq.(\ref{eq10}) can be rewritten in the form of matrix,
\begin{equation}\label{eq11}
{P_j}(\infty ) = \frac{{{Z_{rj}}}}{{\sum\nolimits_{k = 1}^N {{Z_{rk}}} }},
\end{equation}
where we have defined the matrix $\bf{Z}$ as
\begin{equation}\label{eq12}
\bf{Z}=\left( \bf{I}-\bf{\tilde{W}}  \right)^{-1}=\bf{I}+\bf{\tilde{W}}+{\bf{\tilde{W}}}^2+\cdots.
\end{equation}
The entry  $Z_{rj}$ denotes the average time spent on the node $j$ before the particle is reset having started from the resetting node $r$.

\section{Mean first-passage time}\label{sec4}
Let us suppose that there is a trap located at node $j$. Once the particle arrives at the trap, the particle is absorbed immediately. Let us denote by $F_{ij}^{}(t)$ as the probability that the particle visits node $j$ at time $t$ for the first time assuming that the particle has started from node $i$ at $t=0$. The first passage probability $F_{ij}^{}(t)$ and the occupation probability $P_{ij}(t)$ satisfy the following renewal equation  \cite{PhysRevLett.92.118701}, 
\begin{eqnarray}\label{eq13}
{P^{}_{ij}}( t ) = {\delta _{t0}}{\delta _{ij}} + \sum_{t' = 0}^t {{F_{ij}^{}}( {t'} ){P^{}_{jj}}( {t - t'} )} ,
\end{eqnarray}
In the Laplace domain, Eq.(\ref{eq13}) becomes
\begin{eqnarray}\label{eq14}
{{\tilde F^{}}_{ij}}( s ) = \frac{{{{\tilde P^{}}_{ij}}(s) - {\delta _{ij}}}}{{{{\tilde P^{}}_{jj}}(s)}}.
\end{eqnarray}

Furthermore, let us define $Q_{ij}(t)$ as the survival probability of the particle up to time $t$, providing that the particle has started from node $i$ at $t=0$. Obviously,  $F_{ij}^{}(t)=Q_{ij}^{}(t-1)-Q_{ij}^{}(t)$ for $t \ge 1$ and $F_{ij}^{}(0)=0$ for $t =0$. By the Laplace transform, we have ${{\tilde F}_{ij}^{}}\left( s \right) =1+ \left( {{e^{ - s}} - 1} \right){{\tilde Q}_{ij}^{}}( s )$. Therefore, 
\begin{eqnarray}\label{eq15}
\tilde Q_{ij}^{}(s) = \frac{{1 - \tilde F_{ij}^{}( s )}}{{1 - {e^{ - s}}}} = \frac{{\tilde P_{jj}^{}(s) - \tilde P_{ij}^{}(s) + {\delta _{ij}}}}{{\left( {1 - {e^{ - s}}} \right)\tilde P_{jj}^{}( s )}}.
\end{eqnarray}

The MFPT from node $i$ to node $j$ is calculated as $\langle {{T_{ij}}} \rangle = \mathop {\lim }\nolimits_{s \to 0} {\tilde Q_{ij}}( s )$, given by (for Appendix \ref{app3} for details)
\begin{widetext}
\begin{eqnarray}\label{eq16}
\langle {{T_{ij}}} \rangle = \left\{ \begin{array}{lr}
\frac{1}{{{P_j}( \infty  )}}\sum\limits_{\ell = 1}^N {\frac{{\langle { j |{\psi _\ell}} \rangle \langle { {{{\bar \psi }_\ell}} |j} \rangle  - \langle { i |{\psi _\ell}} \rangle \langle { {{{\bar \psi }_\ell}} |j} \rangle }}{{1 - {\lambda _\ell}}} + \sum\limits_{k = 1}^N  } \sum\limits_{\ell = 1}^N {\frac{{ {\langle { i |{\psi _\ell}} \rangle \langle { {{{\bar \psi }_\ell}} |k} \rangle  - \langle { j |{\psi _\ell}} \rangle \langle { {{{\bar \psi }_\ell}} |k} \rangle } }}{{{{ {1 - {\lambda _\ell}} }}}}} , &i \ne j,\\
\frac{1}{{{P_j}\left( \infty  \right)}},&i = j.
\end{array} \right.
\end{eqnarray}
\end{widetext}
Eq.(\ref{eq16}) can be rewritten as the matrix form, 
\begin{eqnarray}\label{eq17}
\langle {{T_{ij}}} \rangle  = \left\{ \begin{array}{lr}
\frac{1}{{{P_j}\left( \infty  \right)}}\left( {{Z_{jj}} - {Z_{ij}}} \right) + \sum\limits_{k = 1}^N {\left( {{Z_{ik}} - {Z_{jk}}} \right)}, &i \ne j, \\
\frac{1}{{{P_j}\left( \infty  \right)}}, & i = j. \\ 
\end{array} \right. \nonumber \\
\end{eqnarray}

It is also useful to quantify the ability of a process to explore
the whole network. For this purpose, we define $T(j)$ as the global MFPT (GMFPT) to the target node $j$ \cite{PhysRevE.80.065104,PhysRevLett.109.088701}, averaged over all the starting node $i$ except for node $j$, 
\begin{eqnarray}\label{eq23}
T( j ) =\frac{1}{N-1} \sum_{i \neq j } {  \langle T_{ij} \rangle } .
\end{eqnarray}
Furthermore, one can average the GMFPT over all nodes and get a property of the whole network which was introduced as the graph MFPT (GrMFPT) \cite{PhysRevE.89.012803},
\begin{eqnarray}\label{eq23_2}
T = \frac{1}{N}\sum_{j } {T( j )} = \frac{1}{N(N-1)} \sum_{j} \sum_{i \neq j}{\langle T_{ij} \rangle} . 
\end{eqnarray}

\section{Node-independent resetting probability}\label{sec5}
For node-independent resetting probabilities, $\gamma_i \equiv \gamma$ for each $i$, Eq.(\ref{eq3}) can be reduced to ${\bf{\tilde W}}=(1-\gamma)  \bf{W}$. Therefore, the eigenvalues of ${\bf{\tilde W}}$, $\lambda_\ell$, and the eigenvalues of ${\bf{ W}}$, $\xi_\ell$, have a simple relation,  $\lambda_\ell=(1-\gamma) \xi_\ell $. Meanwhile, ${\bf{\tilde W}}$ and ${\bf{ W}}$ share the same eigenvectors. Since ${\bf{ W}}$ is a stochastic matrix that satisfies the sum of each row equal to one, its maximal eigenvalue is equal to one. Without loss of generality, we let $\xi_1=1$ and the absolute values of other eigenvalues are always less than one. The right eigenvector corresponding to $\xi_1=1$ is given simply by $| \psi_1 \rangle=(1,\cdots,1)^{\top}$. 

\begin{figure*}
	\centerline{\includegraphics*[width=1.8\columnwidth]{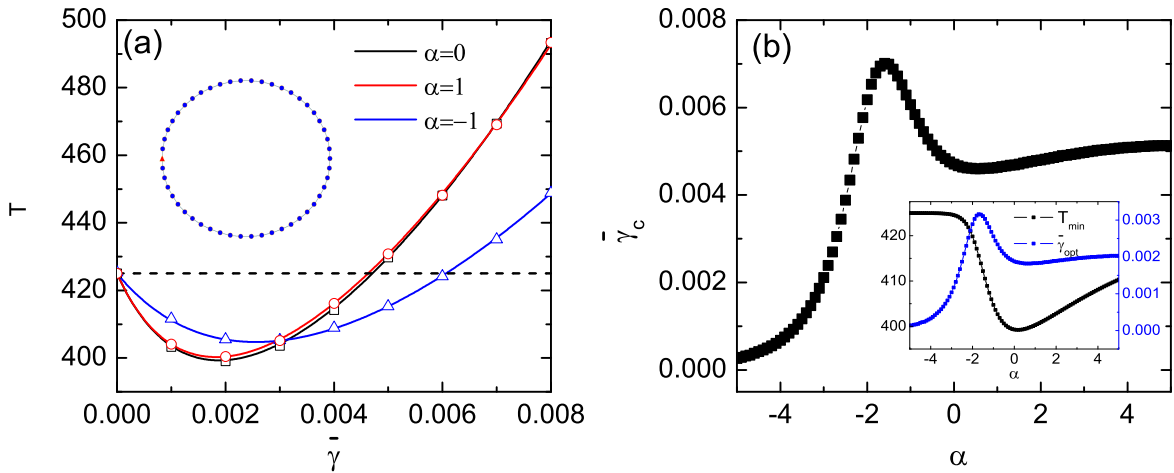}}
	\caption{Results on a ring network of size $N=50$ shown in the inset of (a). $f_i$ in Eq.(\ref{eq19}) is chosen as $f_i=d(i,r)$, where $d(i,r)$ denotes the shortest path length between node $i$ and the resetting node $r$ (red triangle in the inset of (a)).  (a) The GrMFPT as a function of the averaged resetting probability $\bar{\gamma}$ for three different $\alpha$. The horizontal dashed line indicates the result without resetting. Solid lines and symbols represent the theoretical and simulation results, respectively.  (b) $\bar \gamma_{c}$ as a function of $\alpha$. In the inset of (b) we show $T_{\min}$ and $\bar \gamma_{\rm{opt}}$ as a function of $\alpha$.  } \label{fig1}
\end{figure*}

Under such a case, Eq.(\ref{eq10}) can be rewritten as
\begin{eqnarray}\label{eq12.1}
{P_j}(\infty ) &=& \frac{{\sum\nolimits_{\ell = 1}^N {\frac{{\langle r | {{\psi _\ell}} \rangle \langle {{{\bar \psi }_\ell}} | j \rangle }}{{1 - \left( {1 - \gamma } \right){\xi _\ell}}}} }}{{\sum\nolimits_{k = 1}^N {\sum\nolimits_{\ell = 1}^N {\frac{{\langle r | {{\psi _\ell}} \rangle \langle {{{\bar \psi }_\ell}} | k \rangle }}{{1 - \left( {1 - \gamma } \right){\xi _\ell}}}} } }} \nonumber \\ &=& \frac{{\sum\nolimits_{\ell = 1}^N {\frac{{\langle r | {{\psi _\ell}} \rangle \langle {{{\bar \psi }_\ell}} | j \rangle }}{{1 - \left( {1 - \gamma } \right){\xi _\ell}}}} }}{{\sum\nolimits_{\ell = 1}^N {\frac{{\langle r | {{\psi _\ell}} \rangle \langle {{{\bar \psi }_\ell}} | {{\psi _1}} \rangle }}{{1 - \left( {1 - \gamma } \right){\xi _\ell}}}} }} \nonumber \\ &=& \langle {{{\bar \psi }_1}} | j \rangle  + \gamma \sum\nolimits_{\ell = 2}^N {\frac{{\langle r | {{\psi _\ell}} \rangle \langle {{{\bar \psi }_\ell}} | j \rangle }}{{1 - \left( {1 - \gamma } \right){\xi _\ell}}}} .
\end{eqnarray}
In the second line of Eq.(\ref{eq12.1}), we have utilized the facts $|\psi_1 \rangle=\sum_{k=1}^{N} |k \rangle$ and $\langle \bar \psi_\ell | \psi_1 \rangle =\delta_{\ell 1}$. The first term in Eq.(\ref{eq12.1}) is the stationary occupation probability in the absence of resetting \cite{PhysRevLett.92.118701}, and
the second term in Eq.(\ref{eq12.1}) is a nonequilibrium contribution due to the resetting processes.

In the case of a constant resetting probability Eq.(\ref{eq16}) can be rewritten as
\begin{eqnarray}\label{eq18}
\langle {{T_{ij}}} \rangle  =  \left\{ \begin{array}{lr}
\frac{1}{{{P_j}( \infty  )}}\sum\limits_{\ell = 2}^N {\frac{{\langle { j |{\psi _\ell}} \rangle \langle { {{{\bar \psi }_\ell}} |j} \rangle  - \langle { i |{\psi _\ell}} \rangle \langle { {{{\bar \psi }_\ell}} |j} \rangle }}{{1 - (1-\gamma) {\xi _\ell}}}}, &i \neq j,\\
\frac{1}{{{P_j}\left( \infty  \right)}},&i = j.
\end{array} \right.
\end{eqnarray}
Eq.(\ref{eq12.1}) and Eq.(\ref{eq18}) recover to the results of Ref.\cite{PhysRevE.101.062147}. 

\begin{figure*}
	\centerline{\includegraphics*[width=1.8\columnwidth]{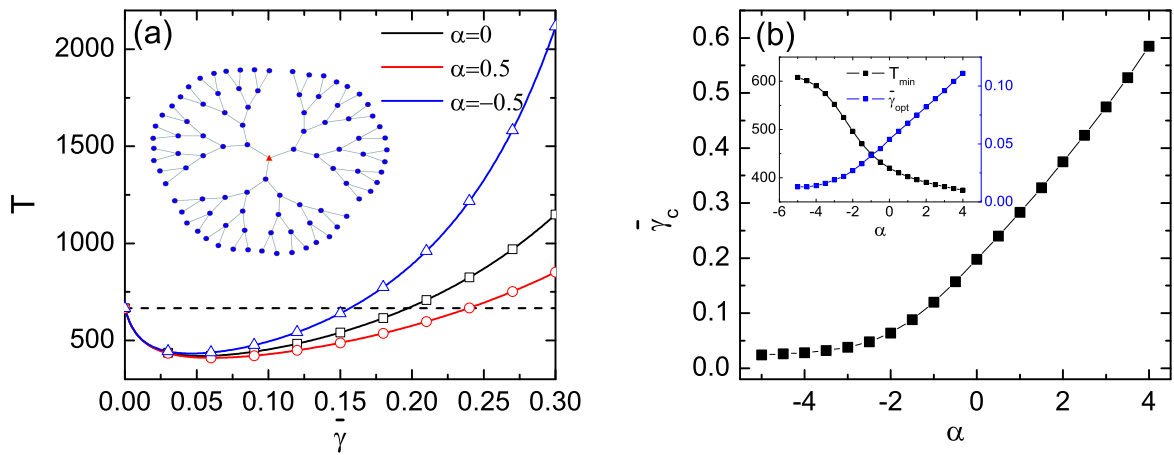}}
	\caption{Results on a finite Cayley tree of size $N=94$ shown in the inset of (a). The resetting node is indicated by red triangle. $f_i$ in Eq.(\ref{eq19}) is chosen as $f_i=d(i,r)$, where $d(i,r)$ denotes the shortest path length between node $i$ and the resetting node $r$ (the root node). (a) The GrMFPT as a function of the averaged resetting probability $\bar{\gamma}$ for three different $\alpha$. The horizontal dashed line indicates the result without resetting. Solid lines and symbols represent the theoretical and simulation results, respectively.  (b) $\bar \gamma_{c}$ as a function of $\alpha$. In the inset of (b) we show $T_{\min}$ and $\bar \gamma_{\rm{opt}}$ as a function of $\alpha$.} \label{fig2}
\end{figure*}

\section{Node-dependent resetting probability}\label{sec6}
As shown in Sec.\ref{sec3} and Sec.\ref{sec4}, we have derived the exact results for the stationary occupation probability and for the MFPT under a general node-dependent resetting probability. We now turn to the specific form of the resetting probability. To the end, we assume that the resetting probability is a function of an attribute of nodes, such as the node's degree, the shortest path length to the resetting node, etc. On the one hand, the assumption is simple enough so that we can conveniently validate our theory. On the other hand, such a consideration may be reasonable from a practical point of view. For example, in the searching process on a network the searcher may collect some local information of its present position, such as node’s degree, etc. Thus, the searcher can adjust its resetting probability in terms of the local information. Since the resetting probability $\gamma_i$ on each node is bounded between 0 and 1, we take $\gamma_i$ as a power function of an attribute $f_i$ of node $i$, subject to an upper limit $\gamma_{\max}$, given by  
\begin{eqnarray}\label{eq19}
{\gamma _i} = \min \left\{ \mu f_i^{\alpha}, \gamma_{\max}  \right\} ,
\end{eqnarray}
where $\alpha$ is a parameter that controls the dependence of resetting probability on node's attribute, and $\mu$ is used to adjust the average value of resetting probabilities. $\gamma_{\max}$ is a cutoff value of resetting probability, and is set to be $\gamma_{\max}=1$ unless otherwise specified. In particular, $\alpha=0$ corresponds to the case of resetting with constant probability \cite{PhysRevE.101.062147}. 

We first consider $f_i=d(i,r)$, where $d(i,r)$ denotes the shortest path length between node $i$ and the resetting node $r$. In Fig.\ref{fig1}, we show the results on a ring network of size $N=50$ (see the inset of Fig.\ref{fig1}(a)), from which we choose one of nodes as the only resetting node. In Fig.\ref{fig1}(a), we plot the GrMFPT as a function of the average resetting probability, $\bar \gamma=N^{-1} \sum_{i=1}^{N} \gamma_i$, for three different values of $\alpha$. We compare the analytical results (solid lines in Fig.\ref{fig1}(a)) against the same obtained from direct numerical simulations (symbols in Fig.\ref{fig1}(a)). In all simulations, we have used
$2 \times 10^3$ averages to estimate the MFPT between arbitrary two nodes. The results are found to be in excellent agreement between theory and simulations. The GrMFPT, $T$, shows a nonmonotonic dependence on $\bar{\gamma}$. There exists an optimal value of $\bar{\gamma}=\bar{\gamma}_{\rm{opt}}$ for which $T$ admits a minimum, $T=T_{\min}$. Comparing to the case without resetting (see horizontal dashed line in Fig.\ref{fig1}(a)), there is a wide range of $\bar{\gamma} \in (0, \bar{\gamma_c})$ for which $T$ can be decreased, in the sense that the resetting is able to optimize the efficiency of searching processes. Obviously, the larger value of $\bar{\gamma_c}$, the wider range for optimizing the GrMFPT comparing with the case without resetting, and
$\bar{\gamma_c}$ is thus a measure for optimization scope via resetting. 

To investigate the impact of the node-dependent protocol on 
$\bar{\gamma_c}$, we calculate $\bar{\gamma_c}$ as a function of $\alpha$, as shown in Fig.\ref{fig1}(b). Also, in the inset of Fig.\ref{fig1}(b), we show $\bar{\gamma}_{\rm{opt}}$ and $T_{\min}$ as a function of $\alpha$. We find that all the three quantities vary nonmonotonically with $\alpha$. Noticeably, $\bar{\gamma_c}$ and $\bar{\gamma}_{\rm{opt}}$ show their maxima at $\alpha=-1.6$, although $T_{\min}$ shows a minimum at $\alpha=0$ (corresponding to the case with a constant probability resetting). This implies that an appropriate negative correlation between the resetting probability at a node and its  distance to the resetting node can expand the scope of optimization
for the GrMPFT on ring networks. This result is counterintuitive because that one may naturally think that it is more beneficial when the resetting happens more frequently in the region away from the resetting node.

\begin{figure*}
	\centerline{\includegraphics*[width=1.8\columnwidth]{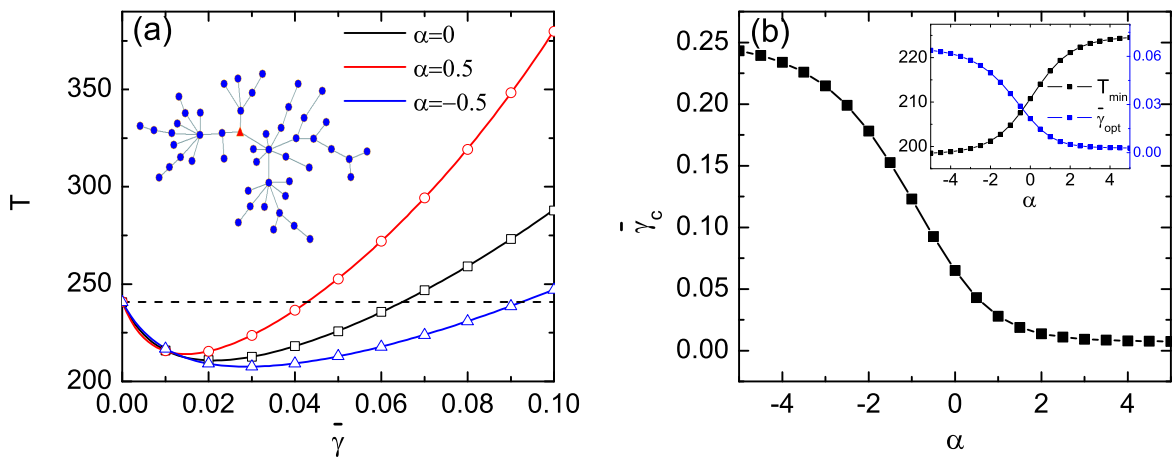}}
	\caption{Results on a BA network of size $N=50$ and average degree $\langle k \rangle=2$ shown in the inset of (a). A node (red triangle) is chosen as the only resetting node. $f_i$ in Eq.(\ref{eq19}) is chosen as $f_i=d_i$, where $d_i$ denotes the degree of node $i$. (a) The GrMFPT as a function of the averaged resetting probability $\bar{\gamma}$ for three different $\alpha$. The horizontal dashed line indicates the result without resetting. Solid lines and symbols represent the theoretical and simulation results, respectively.  (b) $\bar \gamma_{c}$ as a function of $\alpha$. In the inset of (b) we show $T_{\min}$ and $\bar \gamma_{\rm{opt}}$ as a function of $\alpha$.} \label{fig3}
\end{figure*}

In Fig.\ref{fig2}, we show the results on a finite Cayley tree of coordination number
$z=3$ and composed of $n=5$ shells (see the inset of Fig.\ref{fig2}(a)). The nodes in the outermost shell have degree 1, whereas the other nodes have degree $z$. The root node is set to be the only resetting node. In Fig.\ref{fig2}(a), we also observe that the GrMPFT exhibits a minimum at an optimal value of $\bar \gamma_{\rm{opt}}$. Comparing with the case of without resetting (see the horizontal dashed line), the GrMFPT can be accelerated in the range of $0<\bar{\gamma}<\bar{\gamma_c}$. $\bar{\gamma_c}$ shows a monotonic increase with $\alpha$, as shown in Fig.\ref{fig2}(b). This indicates that when the resetting probabilities at the outer nodes are larger than those at the inner nodes, the 
scope of optimization for the GrMFPT becomes wider. Furthermore, as $\alpha$ increases, $\bar{\gamma}_{\rm{opt}}$ shifts to a larger value and $T_{\min}$ is decreased gradually, as shown in the inset of Fig.\ref{fig2}(b).

Finally, we consider the case when the resetting probability depends on the node's degree, i.e., $f_i=d_i$, where $d_i$ is the degree of node $i$. In Fig.\ref{fig3}, we present the result on a Barab\'asi-Albert (BA) network \cite{Science.286.509} of size $N=50$ and average degree $\langle k \rangle=2$ (see the inset of Fig.\ref{fig3}(a)). We choose a node as the only resetting node (red triangle). In Fig.\ref{fig3}(a), we again see that the GrMPFT shows a nonmonotonic change with $\bar{\gamma}$. For $0<\bar{\gamma}<\bar{\gamma_c}$, the GrMPFT is less than that in the absence of resetting (see dashed line in Fig.\ref{fig3}(a)). 
When the resetting probabilities at those nodes with larger degrees are larger than those at those node with smaller degrees, the optimization region shrinks, see for example $\alpha=0.5$ in Fig.\ref{fig3}(a). Conversely, the optimization region is expanded, see for example $\alpha=-0.5$ in Fig.\ref{fig3}(a). In Fig.\ref{fig3}(b), we plot $\bar{\gamma_c}$ as a function of $\alpha$.  $\bar{\gamma_c}$ increases monotonically with $\alpha$. In addition, as $\alpha$ increases, $\bar{\gamma}_{\rm{opt}}$ decreases monotonically and $T_{\min}$ increases slowly, as shown in the inset of Fig.\ref{fig3}(b).

\section{Conclusions}\label{sec7}
To conclude, we have explored the impact of stochastic resetting on the diffusion and first passage properties of discrete-time random walks on networks where the resetting probability is node-dependent. We have derived the exact expressions of stationary occupation probabilities of the walker on each node and the MFPT between arbitrary two nodes. The two quantities (see Eq.(\ref{eq11}) and Eq.(\ref{eq17})) can be calculated from the matrix $\bf{Z}$ defined in Eq.(\ref{eq12}). Our deduction is general and is able to apply any protocol of node-dependent resetting probability. For concreteness we have considered two different resetting protocols on three types of networks. The first resetting protocol under consideration is that the resetting probability is a function of the distance between a node and the resetting node. The other is   dependent on node's degree. To quantify the efficiency of global searching, we have paid our attention to the so-call GrMFPT, that is the MFPT averaged over all pairs of different nodes. The results show that the GrMFPT exhibits a nonmonotonic change with the mean resetting probability $\bar{\gamma}$. There exists a wide range of $\bar{\gamma} \in (0, \bar{\gamma_c}) $ for which the GrMFPT is lower than that in the absence of resetting.  Comparing to the case of constant resetting probability, the scope for optimizing the GrMFPT can be further expanded for certain settings of parameter, thereby embodying the advantage of the node-dependent resetting probability.    

There are still many open questions concerning the resetting paradigm. In this work we only focused on a simple random walk model but one could generalize to other types of random walks, such as biased random walks \cite{PhysRevE.87.012112,PhysRevE.87.062140}, maximum entropy random walks \cite{PhysRevLett.102.160602}, and so on. Moreover, it would be interesting
to consider the effect of resetting costs on searching processes. In this context, how to find an optimal trade-off between minimizing the GrMFPT and the resetting costs is a challenging issue, although some important progress has been made recently in continuous systems  \cite{arXiv:2112.11416}.

\appendix
\section{Spectral decomposition of $\bf{\tilde{W}}$}\label{app1}
Letting ${\bf{U}} = {\left( {{\bf{I}} - {\bf{Y}}} \right)^{1/2}}{{\bf{D}}^{ - 1/2}}$, Eq.(\ref{eq3}) can be rewritten as
\begin{eqnarray}\label{eqa1}
{\bf{\tilde W}} =  {\bf{U}}\left( {{\bf{UAU}}} \right){{\bf{U}}^{ - 1}}= {\bf{U}} \bf{\tilde{A}}{{\bf{U}}^{ - 1}}
\end{eqnarray}
where $\bf{\tilde{A}}={{\bf{UAU}}}$ is a real-valued symmetric matrix that can be expressed in terms of spectral decomposition, 
\begin{eqnarray}\label{eqa3}
{\bf{\tilde A}} = \sum\limits_{\ell = 1}^N {{\lambda _\ell}| {{\phi _\ell}} \rangle } \langle {{\phi _\ell}} |
\end{eqnarray}
where $\lambda_\ell$ is the $\ell$th eigenvalue of $\bf{\tilde A}$, and the corresponding left eigenvector and right eigenvector are respectively $\langle  \phi_\ell |$ and $|\phi_\ell \rangle$, satisfying $\langle  \phi_\ell | \phi_m \rangle =\delta_{\ell m}$, and $\sum_{\ell=1}^{N} |\phi_\ell \rangle \langle \phi_\ell |  =\bf{I}$. In terms of Eq.(\ref{eqa1}), we obtain
\begin{eqnarray}\label{eqa4}
{\bf{\tilde W}}= \sum\limits_{\ell = 1}^{N} {{\lambda_\ell}| {{\psi _\ell}} \rangle } \langle {{{\bar \psi }_\ell}} |
\end{eqnarray}
where the eigenvalues of ${\bf{\tilde W}}$ are the same as those of ${\bf{\tilde A}}$, and eigenvetors of ${\bf{\tilde W}}$ are given by $|{\psi_\ell}\rangle = {\bf{U}}| {\phi _\ell} \rangle$ and $\langle {{{\bar \psi }_\ell}} | = \langle {{\phi _\ell}} |{{\bf{U}}^{ - 1}}$.

\section{Derivation of stationary occupation probability $P_j(\infty)$}\label{app2}
According to Eq.(\ref{eq9}), we have
\begin{eqnarray}\label{eqb1}
{P_{j}}( \infty ) &=& \mathop {\lim }\limits_{s \to 0} \left( {1 - {e^{ - s}}} \right)\tilde P_{ij}^{\rm{nores}}\left( s \right) \nonumber \\ &+& \mathop {\lim }\limits_{s \to 0} \left( {1 - {e^{ - s}}} \right)\frac{{{e^{ - s}}\sum\limits_k {{\gamma _k}\tilde P_{ik}^{\rm{nores}}\left( s \right)\tilde P_{rj}^{\rm{nores}}\left( s \right)} }}{{1 - {e^{ - s}}\sum\limits_k {{\gamma _k}\tilde P_{rk}^{\rm{nores}}\left( s \right)} }} \nonumber \\ &=& \mathop {\lim }\limits_{s \to 0} \left( {1 - {e^{ - s}}} \right)\frac{{{e^{ - s}}\sum\limits_k {{\gamma _k}\tilde P_{ik}^{\rm{nores}}\left( s \right)\tilde P_{rj}^{\rm{nores}}\left( s \right)} }}{{1 - {e^{ - s}}\sum\limits_k {{\gamma _k}\tilde P_{rk}^{\rm{nores}}\left( s \right)} }} \nonumber \\
\end{eqnarray}
In the second line of Eq.(\ref{eqb1}), we have used the fact $\mathop {\lim }\limits_{s \to 0} \left( {1 - {e^{ - s}}} \right)\tilde P_{ij}^{\rm{nores}}( s )=0 $ since all the eigenvalues of $\bf{\tilde{W}}$ are less than one for $\max\{\gamma_1,\cdots\,\gamma_N\}>0$. Furthermore, we turn to evaluate the value of   
${\sum\nolimits_k {{\gamma _k}\tilde P_{rk}^{\rm{nores}}( 0 )} }$. It is not hard to verify
\begin{eqnarray}\label{eqb2}
\left( {{\bf{I}} - {\bf{\tilde W}}} \right)\left( {\begin{array}{*{20}{c}}
	1\\
	\vdots \\
	1
	\end{array}} \right) = \left( {\begin{array}{*{20}{c}}
	{{\gamma _1}}\\
	\vdots \\
	{{\gamma _N}}
	\end{array}} \right)
\end{eqnarray}
As mentioned before, all the eigenvalues of $\bf{\tilde{W}}$ are less than one in the presence of resetting, and thus ${{\bf{I}} - {\bf{\tilde W}}}$ is nonsingular. Taking the inverse of Eq.(\ref{eqb2}), we have
\begin{eqnarray}\label{eqb3}
{\left( {{\bf{I}} - {\bf{\tilde W}}} \right)^{ - 1}}\left( {\begin{array}{*{20}{c}}
	{{\gamma _1}}\\
	\vdots \\
	{{\gamma _N}}
	\end{array}} \right) = \left( {\begin{array}{*{20}{c}}
	1\\
	\vdots \\
	1
	\end{array}} \right)
\end{eqnarray}
or equivalently
\begin{eqnarray}\label{eqb4}
\sum\limits_{k = 1}^N {{\gamma _k}} {\left[ {{{\left( {{\bf{I}} - {\bf{\tilde W}}} \right)}^{ - 1}}} \right]_{ik}} = 1, \quad \forall i
\end{eqnarray}
Eq.(\ref{eqb4}) can be rewritten in the form of spectral decomposition, 
\begin{eqnarray}\label{eqb5}
\sum\limits_{k = 1}^N {{\gamma _k}} \sum\limits_{\ell = 1}^N {\frac{{\langle i | {{\psi _\ell}} \rangle \langle {{{\bar \psi }_\ell}} | k \rangle }}{{1 - {\lambda _\ell}}} = \sum\limits_{k = 1}^N {{\gamma _k}} \tilde P_{ik}^{\rm{nores}}( 0 )}  = 1, \quad \forall i \nonumber \\
\end{eqnarray}
where we have utilized Eq.(\ref{eq5.1}). Therefore, the limit in Eq.(\ref{eqb1}) has the form of 0/0, and thus we then apply the L'H\^opital rule to calculate the limit,
which leads to
\begin{eqnarray}\label{eqb6}
{P_j}( \infty  ) = \frac{{\sum\nolimits_{\ell = 1}^N {\frac{{\langle r | {{\psi _\ell}} \rangle \langle {{{\bar \psi }_\ell}} | j \rangle }}{{1 - {\lambda _\ell}}}} }}{{\sum\nolimits_{k = 1}^N {{\gamma _k}} \sum\nolimits_{\ell = 1}^N {\frac{{\langle r | {{\psi _\ell}} \rangle \langle {{{\bar \psi }_\ell}} | k \rangle }}{{{{\left( {1 - {\lambda _\ell}} \right)}^2}}}} }}
\end{eqnarray}
To simplify Eq.(\ref{eqb6}), we calculate
\begin{eqnarray}\label{eqb7}
{\left( {{\bf{I}} - {\bf{\tilde W}}} \right)^{ - 2}}\left( {\begin{array}{*{20}{c}}
	{{\gamma _1}}\\
	\vdots \\
	{{\gamma _N}}
	\end{array}} \right) = {\left( {{\bf{I}} - {\bf{\tilde W}}} \right)^{ - 1}}\left( {\begin{array}{*{20}{c}}
	1\\
	\vdots \\
	1
	\end{array}} \right)
\end{eqnarray}
where we have used the result of Eq.(\ref{eqb3}). Eq(\ref{eqb7}) can be rewritten in the form of spectral decomposition, 
\begin{eqnarray}\label{eqb8}
\sum\limits_{k = 1}^N {{\gamma _k}} \sum\limits_{\ell = 1}^N {\frac{{\langle i | {{\psi _\ell}} \rangle \langle {{{\bar \psi }_\ell}} | k \rangle }}{{{{\left( {1 - {\lambda _\ell}} \right)}^2}}}}  = \sum\limits_{k = 1}^N {\sum\limits_{\ell = 1}^N {\frac{{\langle i | {{\psi _\ell}} \rangle \langle {{{\bar \psi }_\ell}} | k \rangle }}{{1 - {\lambda _\ell}}}} }, \quad \forall i  \nonumber \\
\end{eqnarray}
Utilizing Eq.(\ref{eqb8}), Eq.(\ref{eqb6}) simplifies to Eq.(\ref{eq10}).

\section{Derivation of the MFPT}\label{app3}
Let $\langle {{T_{ij}}} \rangle$ be the MFPT from node $i$ to node $j$, which can be calculated as
\begin{eqnarray}\label{eqc1}
\left\langle {{T_{ij}}} \right\rangle  &=& \mathop {\lim }\limits_{s \to 0} {\tilde Q_{ij}}\left( s \right) \nonumber \\ &=& \mathop {\lim }\limits_{s \to 0} \frac{{{{\tilde P}_{jj}}\left( s \right) - {{\tilde P}_{ij}}\left( s \right) + {\delta _{ij}}}}{{\left( {1 - {e^{ - s}}} \right){{\tilde P}_{jj}}\left( s \right)}} \nonumber \\ &=& \left\{ \begin{array}{lc}
\frac{1}{{{P_j}\left( \infty  \right)}}\mathop {\lim }\limits_{s \to 0} \left[ {{{\tilde P}_{jj}}\left( s \right) - {{\tilde P}_{ij}}\left( s \right)} \right], & i \ne j \\
\frac{1}{{{P_j}\left( \infty  \right)}}, & i \ne j
\end{array} \right.
\end{eqnarray}
Using Eq.(\ref{eq9}) we calculate the limit,   
\begin{eqnarray}\label{eqc2}
&& \mathop {\lim }\limits_{s \to 0} \left[ {{{\tilde P}_{jj}}\left( s \right) - {{\tilde P}_{ij}}\left( s \right)} \right] \nonumber \\ &=& \mathop {\lim }\limits_{s \to 0} \left[ {\tilde P_{jj}^{\rm{nores}}\left( s \right) - \tilde P_{ij}^{\rm{nores}}\left( s \right)} \right] \nonumber \\ &+& \mathop {\lim }\limits_{s \to 0} \frac{{{e^{ - s}}\tilde P_{rj}^{\rm{nores}}\sum\nolimits_{k = 1}^N {{\gamma _k}\left[ {\tilde P_{jk}^{\rm{nores}}\left( s \right) - \tilde P_{ik}^{\rm{nores}}\left( s \right)} \right]} }}{{1 - {e^{ - s}}\sum\nolimits_{k = 1}^N {{\gamma _k}\tilde P_{rk}^{\rm{nores}}\left( s \right)} }}\nonumber \\
\end{eqnarray}
Substituting Eq.(\ref{eq5.1}) into Eq.(\ref{eqc2}), we can obtain the first term on the r.h.s. of Eq.(\ref{eqc2}). On the other hand, the second term on the r.h.s. of Eq.(\ref{eqc2}) has the form 0/0 in terms of Eq.(\ref{eqb5}), which can be evaluated by the L'H\^opital rule. Finally, we can obtain the MFPT
\begin{widetext}
\begin{eqnarray}\label{eqc3}
\left\langle {{T_{ij}}} \right\rangle  = \left\{ \begin{array}{lr}
\frac{1}{{{P_j}\left( \infty  \right)}}\sum\limits_{\ell = 1}^N {\frac{{\langle { j |{\psi _\ell}} \rangle \langle { {{{\bar \psi }_\ell}} |j} \rangle  - \langle { i |{\psi _\ell}} \rangle \langle { {{{\bar \psi }_\ell}} |j} \rangle }}{{1 - {\lambda _\ell}}} + \sum\limits_{k = 1}^N {{\gamma _k}} } \sum\limits_{\ell = 1}^N {\frac{{{\lambda _\ell}\left( {\langle { i |{\psi _\ell}} \rangle \langle { {{{\bar \psi }_\ell}} |k} \rangle  - \langle { j |{\psi _\ell}} \rangle \langle { {{{\bar \psi }_\ell}} |k} \rangle } \right)}}{{{{\left( {1 - {\lambda _\ell}} \right)}^2}}}} , &i \ne j\\
\frac{1}{{{P_j}\left( \infty  \right)}},&i \ne j
\end{array} \right.
\end{eqnarray}
\end{widetext}
Utilizing Eq.(\ref{eqb5}) and Eq.(\ref{eqb8}), Eq.(\ref{eq18}) simplifies to Eq.(\ref{eq16}).

\begin{acknowledgments}
	This work is supported by the National Natural Science Foundation of China (Grants No. 11875069, No 61973001).
\end{acknowledgments}


\end{document}